\documentclass[runningheads]{llncs}
\usepackage{amsfonts}
\usepackage{amsmath}
\usepackage{booktabs} 
\usepackage{graphicx}
\usepackage{pgf,tikz,hyperref}
\usepackage[all]{nowidow}
\usepackage[utf8]{inputenc}
\usepackage{bigdelim}
\usepackage{amssymb}
\usepackage{color, colortbl}
\newcommand*\xor{\mathbin{\oplus}}
\newcommand{\MyLQ}{$l_\mathbf{q}\:$}

\DeclareMathOperator{\trs}{\mathbf{t}}
\DeclareMathOperator{\dtrs}{\tilde{\trs}}
\DeclareMathOperator{\ltrs}{\trs^\prime}
\DeclareMathOperator{\classy}{\mathcal{C}_\delta^\theta}
\definecolor{Gray}{gray}{0.9}

\graphicspath{{./figures/}}
\DeclareGraphicsExtensions{.pdf,.png}

\definecolor{lime}{HTML}{A6CE39}
\DeclareRobustCommand{\orcidicon}{%
\begin{tikzpicture}
\draw[lime, fill=lime] (0,0)
circle [radius=0.16]
node[white] {{\fontfamily{qag}\selectfont \tiny ID}};
\draw[white, fill=white] (-0.0625,0.095)
circle [radius=0.007];
\end{tikzpicture}
\hspace{-2mm}
}

\foreach \x in {A, ..., Z}{%
\expandafter\xdef\csname orcid\x\endcsname{\noexpand\href{https://orcid.org/\csname orcidauthor\x\endcsname}{\noexpand\orcidicon}}
}

\begin{document}
\title{Playing with blocks}
\subtitle{Toward re-usable deep learning models for side-channel profiled attacks}

%
%
\author{Servio Paguada\inst{1,2}\orcidA{} \and
Lejla Batina\inst{1}\orcidB{} \and
Ileana Buhan\inst{1} \and 
Igor Armendariz\inst{1}\orcidD{}}

\authorrunning{Paguada \textit{et al.}}
%
\institute{Radboud University, Nijmegen, The Netherlands
\email{\{servio.paguadaisaula,lejla.batina,ileana.buhan\}@ru.nl}\\ \and
IKERLAN Research Center, Mondragon/Arrasate, Spain\\
\email{\{slpaguada,iarmendariz\}@ikerlan.es}}

\maketitle              
\begin{abstract}
This paper introduces a deep learning modular network for side-channel analysis. Our deep learning approach features the capability to exchange part of it (modules) with others networks. We aim to introduce reusable trained modules into side-channel analysis instead of building architectures for each evaluation, reducing the body of work when conducting those. Our experiments demonstrate that our architecture feasibly assesses a side-channel evaluation suggesting that learning transferability is possible with the network we propose in this paper.

\end{abstract}
\section{Introduction}
\label{sec:intro}
In the side-channel analysis (SCA) field of research, deep learning models (DL models) are powerful tools to evaluate the implementation of secure algorithms. Unfortunately, despite the significant accomplishments by using deep learning models, many challenges remains.

When evaluating a secure implementation of an IoT device, for example, it is challenging to develop a deep learning classifier that feasibly assesses the resilience of the devices. Electronic noise as countermeasure and desynchronization are specific challenges during the evaluation. Indeed, a noisy signal intrinsically suggests dealing with high-dimensional signals. For instance, targeting a modern System-on-Chip with high clock frequencies requires increasing the sampling resolution; as consequence, the side-channel information required for the evaluation contains leakage traces with several irrelevant features (sample points). Then, noise filters and feature engineering as pre-processing steps are being reconsidered as tools to deal with those challenges~\cite{picek2017secrets,paguada2021towardAE,wu2020remove,Mukhtar2020MachineLearning,mangard2008power,Kong2018OnFS}.

This paper proposes a new technique to overcome those challenges and introduces a novel approach that uses a deep learning classifier whose part of its architecture allows to be re-used in other models that use the same approach. By featuring the exchangeable modules, we can re-use networks for different SCA evaluations, reducing the body of work of deriving models each time. The suggested architecture comprises \textit{coupled} modules, and those modules have specific tasks to deal with the challenges of an SCA evaluation. We call this approach DL-SCA modular network. Precisely, an autoencoder and a convolution base classifier are the two modules that we suggest in this paper.

An autoencoder can effectively deal with the problem of high dimensionality and the problem of noise. An autoencoder comprises two parts by itself; the encoder and decoder. The encoder ends to an embedding where high dimensional leakage traces are transformed into lower dimensional version of them. Because of it, autoencoder are learning algorithms used in pre-processing steps \textit{i.e} feature extraction~\cite{Mukhtar2020ImprovedHA,paguada2021towardAE,Choudary2018EPortableTA}.

The classifier module serves two objctives; (i) the classification required for the SCA evaluation and (ii) to regularize the autoencoder. As we explain in further sections of this paper, autoencoders might fail to compress the samples taken from the device under test; so penalizing it with a regularization might correct it toward better performance.

Our experiment uses datasets with desyncronization and countermeasures. After proving the effectiveness of the DL-SCA modular network, we perform a second set of experiments where we exchange the modules between modular networks. Our results show that transferability is feasible and applicable to side-channel analysis. The contributions of this paper are as follow:

\begin{itemize}
\item We introduce an approach called DL-SCA modular network and deep learning architecture featuring the exchange of modules through models. We provide the implementation details of the architecture, as well as the hyperparameter to take into account in the design to avoid pitfalls.
\item We present a training strategy based on sharing weight technique and early stopping policy for a seamlessly adoption of our approach in current SCA evaluations.
\item We elaborate experiments that demonstrate the effectiveness of re-using modules through modular networks, using different "sharing" protocols based on non-trainable layers.
\end{itemize}

The rest of the paper is organized as follows:
Sect.~\ref{sec:background} details theoretical aspects of the topics used for this work. Related works are discussed in Sect.~\ref{sec:related_works}. Sect.~\ref{sec:datasets} provides information about datasets used in the experiments. Sect.~\ref{sec:dl-sca_modular_network} discuss the main contribution of this paper. Sect.~\ref{sec:training_modules_exp_results} and Sect.~\ref{sec:module_re-usability_exp_results} discuss the experiments. While, Sect.~\ref{sec:conclutions} concludes the paper.

\section{Background}
\label{sec:background}

\subsection{Profiled attack}
A side-channel attack requires a \textit{leakage model} to attack the sensitive information contained in a target device. A leakage model refers to a function ($\delta$) that models the leak of sensitive information. Using a leakage model, an adversary can steal the secret key from a device that implements a cryptographic algorithm. The expression~(\ref{eq:leakage_model}) is an example of leakage model used to attack a cryptographic implementation of AES\footnote{Or any other cryptographic primitive with non-linear functions}. In that leakage model, $p$ is the publicly available data \textit{i.e.} the plaintext and $k^*$ is the secret key.

\begin{equation}
\delta = \text{S-box}(p \xor k^*)
\label{eq:leakage_model}
\end{equation}

The adversary measures the power consumption\footnote{Being the most traditional measurement used in SCA, others are Electromagnetic Emanation, heat, sound, and a few more.} when the device inputs the AES algorithm with $p$ random values from the keyspace $\mathcal{K}=\{0,\cdots 255\}$, drawing several leakage traces (a.k.a power traces). With enough leakage traces, the adversary can find a correlation between the power consumption and the inputs $p$ of the leakage model; consequently, he can infer the key $k^*$.

The previous paragraph just describes a traditional non-profiled side-channel attack over a crypto primitive. However, to understand how a profiled side-channel attack works we have to explain its differences from a non-profiled attack. A profiled type of attack born with the idea of training classifiers to distinguish the outputs of a leakage model; so that the attack splits into two phases; (i) to train a classifier (profiling phase) and (ii) to perform the attack (attack phase).

The first phase comprises applying the corresponding leakage model to a clone device (a.k.a. profile device), collecting from it the leakage traces forming a set of profiling traces ($\mathcal{X}$) used to train the classifier. During the attack phase, a set of attacks traces from the actual device is collected and used for the trained classifier to compute probabilities, then a key recovery process takes place using an algorithm called guessing entropy that we will explain in brief.

Template attack and machine learning are two techniques to build a classifier to evaluate side-channel attacks~\cite{Choudary2018EPortableTA,Lerman2017}. It is well-known that to come up with a classifier for SCA evaluation is not straightforward. Indeed, to reduce the body work of building classifier anytime an SCA evaluation is required is the motivation for the proposal of this paper. We propose a deep learning-based model whose architecture allows the model to exchange classifiers with other deep learning model aimed to conduct SCA evaluation over a different device; and in the following, we address the necessary aspects that support the theory about this approach.

\subsection{Guessing entropy (GE)}
\label{sec:metric}
GE is the average \textit{rank} of the correct key byte value $k^*$ in a key guessing vector $\mathbf{g}$, over all the set $\mathcal{K}$ of key candidates $k$~\cite{Standaert2009UnifiedFramework}.
Formally denoted as $\textrm{GE} = rank_{k^*}(\mathbf{g})$, where $rank_{k}(\mathbf{g}) \in \{0, \dots, |\mathcal{K}| - 1\}$, and the key guessing vector is defined as: $\mathbf{g} = sort(\mathbf{E}[log \mathbf{P_r}(\trs_i; \classy)])$. $\mathbf{P_r}(\trs_i; \classy)$ is the input vector of probabilities $\mathbf{p}_{i,j}$ from a classifier (usually aimed for key recovery task) given a leakage trace $\trs_i$. After applying the expectation $\mathbf{E}$ per multiples experiments of $\mathbf{P_r}$, the \textit{sort} function orders the resultant vector $\mathbf{g}$ in decreasing order. The element $g_0\in\mathbf{g}$ corresponds to the most likely key candidate, while $g_{|\mathcal{K}| - 1}\in\mathbf{g}$ is the less likely one.

\subsection{Deep learning base profiled attacks} 
A deep learning classifier outputs a vector of probabilities fed into the guessing entropy (GE) metric to compute the rank of the key ($k^*$). We denote a deep learning model $\classy$ for profiled attacks as a classifier $\mathcal{C}$ with a vector of parameters $\theta \in \mathbb{R}^n$ aimed to distinguish leakage traces labeled using a leakage model $\delta$. Having labeled leakage traces means that our learning approach is supervised learning~\cite{goodfellow2016deep} which represents one of the most feasible ways to leverage the learning of a deep learning classifier. 

Despite several deep learning architectures, CNNs based models are the preferred architecture to use in profiled attacks. The convolutional part plays an essential role when leakage traces are desynchronized. The deep learning model we propose uses a specific type of convolution, called dilated convolution~\cite{paguada2020ForgottenHy} for boosting the feature extraction capability of the layer (see sub-section~\ref{sec:autoencoder}).

\subsection{Feature extraction}
A feature extraction process applies a transformation (linear or non-linear) to a space of observations resulting in a new space mapped by the transformation. Formally, given profiling set $\mathcal{X}$ of $N$ leakage traces and each trace comprises $m$ features (or sample points). Feature extraction applies a function $\digamma$ to the profiling set $\mathcal{X}$ mapping a new profiling set $\mathcal{Y}$ whose elements have fewer dimensions of the corresponding elements in $\mathcal{X}$; precisely, $\digamma$ is an application such as $\digamma\colon\mathcal{X} \longmapsto \mathcal{Y}$, and $\mathcal{X} \in \mathbb{R}^m,\: \mathcal{Y} \in \mathbb{R}^n$ such that $n<m$.

This transformation aims to derive new features ($\mathcal{Y}$) to leverage the performance of a classifier, for instance. Theoretically, features in $\mathcal{Y}$ contain the ``transformed'' information that best represents the ground truth of $\mathcal{X}$, in the SCA case, it is the leakage of the sensitive information. In simple words, the intensity of the valuable information gets emphasized while the irrelevant information (non-correlated information) has little to no influence in the new space.

However, it is not straightforward to come up with a transformation $\digamma$ that indeed emphasized the side-channel information. A transformation that goes wrong discards a lot of useful information, and it happens when $\digamma$ cannot keep the variance that distinguishes a leakage trace from another; as consequence, $\mathcal{Y}$ is made of several collapsed traces becoming useless for classification purposes. In section~\ref{sec:dl-sca_modular_network}, we will discuss how our proposed method implements regularization to avoid transformations that collapse the $\mathcal{Y}$ space.

Function $\digamma$ can be inferred directly from $\mathcal{X}$. For instance, Principal Components Analysis (PCA)~\cite{dunteman1989principal} or Linear Discriminant Analysis (LDA)~\cite{blei2003latent} are two algorithms to build linear base $\digamma$ functions for feature extraction. However, PCA and LDA are highly sensitive to desynchronization because of their ``per feature'' process, meaning they find a relation by correlating the same positioning feature through samples. So that, when the samples have a spatial disruption, the relation gets reduced, requiring more samples.

\subsection{Autoencoders}
\label{sec:autoencoder}
An autoencoder is a learning algorithm useful to infer $\digamma$; contrary to PCA and LDA, an autoencoder can infer a non-linear transformation due to the non-linear activation functions in its architecture. Moreover, when the autoencoder architecture comprises convolution layers, it handles the spatial disruption better than PCA and LDA.

An autoencoder consists of two parts; (i) an encoder $\varphi$ and (ii) a decoder $\psi$. Let us define a leakage trace $\trs_i \in \mathcal{X}$, and its dimension being denoted as $dim(\trs_i) = m$. The encoder outputs a new trace $\ltrs_i$ with $dim(\ltrs_i) < dim(\trs_i)$ (see expression~(\ref{eq:encoder_decoder})). At the other side of the autoencoder, the decoder tries to reconstruct $\trs_i$ but it is able to re-build an approximation $\dtrs_i$ only; consequently, one can understand that an autoencoder learns by minimizing the difference between $\trs_i$ and $\dtrs_i$ (as we will see in expression~\ref{eq:mean_squared_error}).


\begin{equation}
\ltrs_i = \varphi(\trs_i), \quad \dtrs_i = \psi(\ltrs_i)
\label{eq:encoder_decoder}
\end{equation}

\begin{figure}[!ht]
\centering
\includegraphics[width=0.98\textwidth]{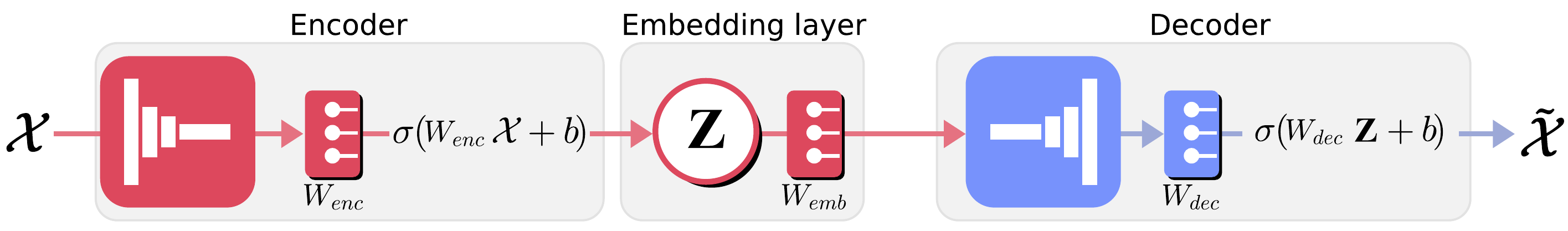}
\caption{Typically autoencoders are symmetrical models, meaning that both parts of the encoder and decoder resemble each other. During training, the encoder trains to code the original signal to a \textit{Latent space}, ideally this code from the features that better represent the characteristic of the original signal. From there, the decoder re-constructs as much as possible the original signal.}
\label{fig:flat_autoencoder_shadows}
\end{figure}

From a functional perspective, the encoder maps $\mathcal{X}$ to an embedding space denoted by $\mathbf{Z}$ (i.e. $\varphi\colon\mathcal{X} \mapsto \mathbf{Z}$), the embedding $\mathbf{Z}$ is usually called latent space, code, latent code or hidden code. Further, $\mathbf{Z}$ is the space result of the transformation applied by the encoder. According to the discussion in the previous sub-section, $\mathbf{Z}$ is the resultant space of a feature extraction process \textit{i.e.} $\mathcal{Y}$. Likewise, the decoder maps $\mathbf{Z}$ to $\tilde{\mathcal{X}}$ (\textit{i.e.} $\psi\colon\mathbf{Z} \mapsto \tilde{\mathcal{X}}$), where $\dtrs \in \tilde{\mathcal{X}}$. The expressions~(\ref{eq:functional_ae_a}) and~(\ref{eq:functional_ae_b}) formalize these two mappings;

\begin{eqnarray}
\label{eq:functional_ae_block}
\mathbf{Z} = \varphi(\mathcal{X})
& = & \sigma(W_{enc}\,\mathcal{X}+b)
\label{eq:functional_ae_a}\\
\tilde{\mathcal{X}} = \psi(\mathbf{Z})
& = & \sigma(W_{dec}\,\mathbf{Z}+b^\prime)
\label{eq:functional_ae_b}
\end{eqnarray}
Function $\sigma$ denoted a non-linear activation function. An encoder is parameterized by a weight matrix $W_{enc} \in \mathbb{R}^{m \times n}$ and a bias vector $b \in \mathbb{R}^n$; likewise, a decoder is parameterized by a weight matrix $W_{dec} \in \mathbb{R}^{n \times m}$ and a bias vector $b^\prime \in \mathbb{R}^m$ (see Fig.~\ref{fig:flat_autoencoder_shadows}).
Training an autoencoder implies finding a vector of parameters $\theta=(W_{enc}, W_{dec},b, b^\prime)$ that minimize a loss function $\mathcal{L}$ such as; 
\begin{equation}\label{eq:loss_theta}
\Theta=\min_\theta \mathcal{L}(\trs, \dtrs) = \min_\theta \mathcal{L}(\trs,\psi(\varphi(\trs)))
\end{equation}

As we said, autoencoder learns by minimizing the difference between $\trs_i$ and $\dtrs_i$; so that, the Mean Square Error (MSE) is a loss function commonly used;
\begin{equation}\label{eq:mean_squared_error}
\mathcal{L}_{\textrm{MSE}} = \mathcal{L}(\trs,\dtrs) = \frac{1}{m} \sum^{m}_{i=1}{(\trs[i] - \dtrs[i])^2}
\end{equation}

\paragraph{Convolution layer architecture} Autoencoders are built using either fully connected layers or convolutional layers. The latter makes the autoencoder inherit the spatial invariant robustness property, which is useful when leakage traces are desynchronized; our autoencoder uses dilated convolution layers.

A convolution layer consists of kernels that essentially are matrices; then, to dilate a convolution layer consists of inserting zeros into its kernels, meaning to separate the matrices' elements using zeros, expanding their receptive field\footnote{Called it to those kernel elements which are not zero}. According to~\cite{paguada2020ForgottenHy} a dilated kernel allows convolutions base classifiers to combine spread features that contain the leakage information, at the same time, avoiding irrelevant features that might lay in between.

Let us consider the expression in (\ref{eq:normal_convolution}) showing a regular convolution where a leakage trace $\trs_i$ is multiplied by the kernel $\mathbf{q}$ whose length is denoted by $l_\mathbf{q}$. If we displace the leakage trace $\trs_i$ from right to left, a single feature $t$ of $\trs_i$ is multiplied $l_\mathbf{q}$ times. If $l_\mathbf{q}$ is large, then $t$ might be \textit{excessively} used during the operation. According to~\cite{Zaid2019}, this excessive use of $t$ may decrease the convolution effectiveness. Notice that if $l_\mathbf{q}$ increases aiming to use further spread features, it also increases the times $t$ is used. By using dilated convolutions, one can avoid this downside.

\setlength{\arraycolsep}{0.0em}
\begin{eqnarray}\label{eq:normal_convolution}
\nonumber
(\trs \circledast\:\mathbf{q})[t]&{}={}&\sum^\infty_{n=-\infty}{\trs[t-n] \cdot \mathbf{q}[t]} \\\nonumber
&{}={}& \cdots + \\
&&\underbrace{(\trs[t-n_i] \cdot \mathbf{q}[n_i]) + (\trs[t-n_{i+1}] \cdot \mathbf{q}[n_{i+1}])}_\text{\MyLQ times} \\\nonumber
&& + \cdots 
\end{eqnarray}
\setlength{\arraycolsep}{5pt}

The expression in (\ref{eq:dilated_convolution}) shows a dilated kernel with one zero inserted between its elements. Notice that when the convolution is performed, the feature $t$ alternates being multiplied or not by a zero; consequently, it reduces the times the operation uses the feature $t$.

\begin{eqnarray}
\nonumber\label{eq:dilated_convolution}
(\trs \circledast\:\mathbf{q}_d)[t]&{}={}&\sum^\infty_{n=-\infty}{\trs[t-n] \cdot \mathbf{q}_d[t]} \\\nonumber
&{}={}& \cdots + \\
&&\left. \begin{array}{l}
(\trs[t-n_i] \cdot 0) + \\
(\trs[t-n_{i+1}] \cdot \mathbf{q}_d[n_{i+1}]) + \\
(\trs[t-n_{i+2}] \cdot 0) + \\
(\trs[t-n_{i+3}] \cdot \mathbf{q}_d[n_{i+3}])
\end{array}
\, \right\} \hat{l_{\mathbf{q}}}\: \text{times}\\\nonumber
&& + \cdots
\end{eqnarray}
The hyperparameter~\textit{dilatation rate} ($dr$) controls the number of zeros inserted. When a kernel is dilated its receptive field is modified by the relation;

\begin{equation}
\hat{l_{\mathbf{q}}} = {l_{\mathbf{q}}+(l_{\mathbf{q}}-1)(dr-1)}
\label{eq:dilated_conv_arithmetic}
\end{equation}

In this way, the receptive field increases by modifying either the length of the kernel or the dilatation rate, letting the user regularize the convolution operation.

\section{Related work}
\label{sec:related_works}
While few works in SCA discuss an approach of architecture transferability with reusable modules, several works have discussed feature reduction for SCA. Cagli~\textit{et al}. in~\cite{cagli2016kernel,cagli2015enhancing,cagli2018feature} discussed application of traditional feature reduction methods using PCA~\cite{dunteman1989principal}, LDA~\cite{blei2003latent}, and its kernel base variant Kernel PCA and KDA. Picek~\textit{et al}.~\cite{picek2019systematic} published results using same methods as~\cite{cagli2015enhancing}. However, authors in~\cite{picek2019systematic} used an approach that combined feature extraction and feature selection; precisely, PCA and LDA combined with SOST and SOSD, they called it hybrid feature selection methods.

Intrinsically, any work that uses the same feature reduction techniques aims to downsample the signal by taking it to a new space (latent space). However, these approaches consider only linear base feature reduction disregarding the more powerful non-linear version of it; it is likely, that this situation may be a consequence of advertising CNNs as built-in feature extraction deep learning models. Hence, very few works have addressed non-linear methods for SCA evaluation. One of those few works are, for instance, Paguada \textit{et al}.~\cite{paguada2021towardAE}, and Yang \textit{et al}.~\cite{yang2020cdae}; similar to us, those works used autoencoders toward inferring a non-linear function to pre-process leakage traces in a fashion that overcome linear methods.

\textit{While those two works are the closest one we can relate with, to our best knowledge, there is no previous work on side-channel analysis that suggest a deep learning approach based on modules; featuring to share modules between models.}

\section{ASCAD fixed and random datasets}
\label{sec:datasets}
ASCAD dataset\footnote{This dataset is publicly available at https://github.com/ANSSI-FR/ASCAD} was introduced in~\cite{Prouff2018StudyOD}. The leakage traces were collected from an Atmega8515 8-bit microcontroller. The cryptographic algorithm implemented is AES-128 protected using masking countermeasure~\cite{joan2002DesignRijndael,Bloemer2004}. 

The dataset has two versions, traces collected with fixed key encryption $k_f$ and traces collected with random key encryption $k_r$ (plaintext is always random), while the target byte of the secret key in both cases is the third one. We named these versions as $\text{ASCAD}^f$ and $\text{ASCAD}^r$ respectively. Due to these key characteristics, $\text{ASCAD}^r$ is more challenging and more realistic than $\text{ASCAD}^f$ when conducting an SCA evaluation over them. TABLE~\ref{tab:ascad_fixed_random_summary} contains a summary of main characteristics of these two datasets.

\begin{table}[!ht]
\renewcommand{\arraystretch}{1.3}
\newcolumntype{a}{>{\columncolor{Gray}}l}
\centering
\begin{tabular}{l l a a}
\hline
\multicolumn{2}{c}{\bfseries $\text{ASCAD}^f$} & \multicolumn{2}{c}{\cellcolor{Gray}\textbf{$\text{ASCAD}^r$}}\\
\hline
Profiling\_traces & $50\,000$ & Profiling\_traces & $200\,000$\\
Attack\_traces & $10\,000$ & Attack\_traces & $100\,000$\\
dim($\trs_i$) & $700$ & dim($\trs_i$) & $1\,400$\\
\hline
\end{tabular}
\caption{Cardinalities of the ASCAD datasets. Since their goal is to be used for benchmarking profiled attacks the leakage traces are grouped in profiling\_traces and the attack\_traces sets.}
\label{tab:ascad_fixed_random_summary}
\end{table}
Leakage traces in each version are desynchronized according to a threshold value that moves traces around the $x$-axis, being frequently used threshold values of 0, 50, and 100. Then, to make clear distinctions when exchanging the modules between modular networks, we add to the name the threshold value, for instance, $\text{ASCAD}^r\:desync50$.

\section{DL-SCA modular network architecture}
\label{sec:dl-sca_modular_network}
This section explains the details about the architecture of the DL-SCA modular network; further, we describe the strategy to train it.

\begin{figure}[!ht]
\centering
\includegraphics[width=0.98\textwidth]{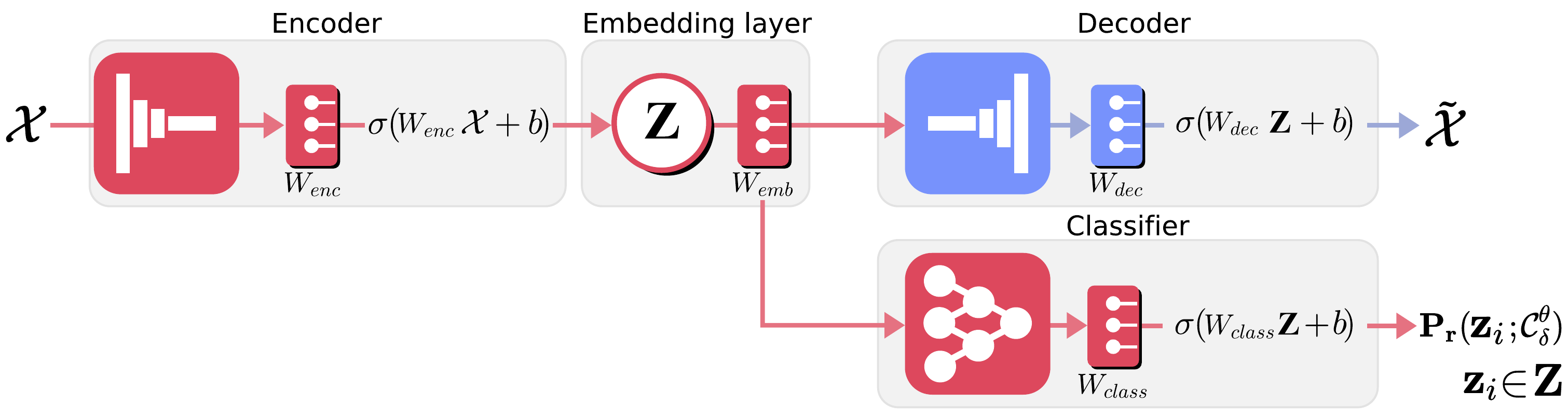}
\caption{DL-SCA modular network architecture illustration. The encoder, with its embedding layer and the classifier ensemble the final model used to perform the attack.}
\label{fig:SCA-CateAE_illustration}
\end{figure}

Since we are using autoencoders; then, our suggested DL-SCA modular network comprises three main modules; an encoder, a decoder, and a classifier (see Fig.~\ref{fig:SCA-CateAE_illustration}). Particularly, we will group the encoder and decoder into a single module called a \textit{downsampler}. The downsampler has two goals; (i) to extract meaningful features by reducing the noise in the leakage traces and (ii) to downsample them. Now, the classifier is in charge of evaluating those extracted features as a classification problem.

It is worth mentioning that once the DL-SCA modular network is trained, we discard the decoder of the downsampler, and we only use the encoder and classifier to perform the SCA evaluation. Due to this, we elaborate a \textit{training strategy} to monitor only those two parts of the model; we will elaborate this late in this section.

The goal of both modules might be apparent; however, the downsampler has an implicit objective. To achieve compatibility with as many classifiers as possible, we should use a downsampler to fix the classifier input. Precisely, we downsample the leakage traces to a fixed length; then, when we re-use the classifier with another downsampler, this latter fixes its output to match the classifier's input. By doing this, we fulfill the first step of re-usability. We demonstrate this in the experimental section of this paper.

Training a DL-SCA modular network architecture requires a loss function for the decoder and another for the classifier. The decoder's loss function ($\mathcal{L}_{MSE}$) was discussed in the sub-section~\ref{sec:autoencoder}. We introduce the classifier loss function.

\subsection{Classifier loss function}

As we said, a classifier outputs a vector of probabilities used as input for the guessing entropy. So, for the classifier to output this vector, it must be trained using a cross-entropy ($CE$) loss function. In supervised profiled side-channel attacks the leakage traces are labeled by the output of a leakage model (see expression (\ref{eq:leakage_model})). Further, the classifier learns by minimizing its error in predicting the label of each trace.

To explain this better, let us consider the expression (\ref{eq:cat_loss_function}). The space $\hat{\mathcal{K}}$ corresponds to a batch of key candidates or labels, each one of the labels in $\hat{\mathcal{K}}$ represents a trace. For instance, let us take $\delta_i \in \hat{\mathcal{K}}$ as one of those labels, we say that $\delta_i$ is the ground truth while $\sigma(\delta_i)$ is the output score a neural network computed\footnote{Often $\sigma$ is the softmax activation function for multi-class classification}.
\begin{equation}
\mathcal{L}_{CE} = -\sum_{i}^{\hat{\mathcal{K}}}\delta_i\cdot \log(\sigma(\delta_i))
\label{eq:cat_loss_function}
\end{equation}
During training, this loss function computes the error in the prediction made by the classifier; consequently, the weights of the classifier are updated toward achieving a prediction with highest accuracy possible.

Clearly, we use a classifier with the same purpose as in a common profiled side-channel evaluation. However, the feature in using a classifier in our approach is to add a regularization term to the downsampler. Precisely, the supervised classifier adds an extra penalization to the downsampler with regard to the featuring space $\mathcal{Y}$ the downsampler is building up; leading the whole network toward better performance. It is impossible to feature this with a self-supervised downsampler trained separately as authors did in~\cite{paguada2021towardAE}.

The arrangement depicted in Fig.~\ref{fig:SCA-CateAE_illustration} suggests that both the classifier and decoder attach to the encoder. Consequently, when training the modular network, the classifier feed-forwards the downsampled traces from the embedding and back-forwards its loss. Meanwhile, the decoder trains its reconstruction capability that additionally penalizes the encoder. These two losses resemble a double voting system that the encoder uses to leverage learning.

Now, notice that because the activation functions are non-linear, the classifier acts as a non-linear regularizer for the embedding space. Consequently, the decoder takes the regularization effect as small perturbations in that space; those perturbations challenge the decoder in reconstructing the original traces as it understands that those are small errors in its reconstruction. Contrary to the approach in~\cite{paguada2021towardAE}, training jointly the autoencoder and the classifier produces an embedding likely to learn positive features. Due to the regularization factor, less correlated features are emphasized over highly correlated noisy features.

\subsection{Analogy with linear regularized autoencoder} 
Autoencoders aim to be imperfect models; so, when training an autoencoder we must avoid an architecture that ends with a model called ``identity function''. When this phenomenon happens, the autoencoder will just copy the data from the input to the output. One way to avoid this is by using a undercomplete architecture, which refers to the embedding we discussed early; further, the deepest an autoencoder is the stronger becomes to avoid ending as an identity function. However, doing this carelessly might reduce the network performance as the model becomes supra-complex, so that, we cannot rely on it repeatedly.

Applying a regularizer to the latent space is another alternative. Regularized autoencoder proved overcoming normal autoencoders when leveraging meaningful features in the embedding. A linear regularizer applies to the latent space an extra penalization. The embedding neurons fire the additional penalization to the decoder added to its loss function as small epsilons of error. This latter advocates the disruption by training its neurons to reconstruct the original data; ignoring that it is being fooled by the regularizer, so its learning is actually ``imperfect''~\cite{goodfellow2016deep}.

A drawback of using linear regularizer is precisely its nature. A linear regularizer applies the penalization linearly to all the embedding neurons, there is no a criterion that controls the magnitude each neuron should receive based on its contribution to the loss function; as eventually, the decoder starts copying the input as it is.

In a DL-SCA modular network, the classifier acts as a regularizer; nonetheless, the regularization is based on non-linearity since the classifier is a non-linear function. The non-linear activation functions used in the classifier receive their input from the embedding neurons; once the classifier does the back-propagation it applies an epsilon value according to their contribution to the classification. Once again the decoder interprets those as small errors, but now facing a more advance regularization.

Both linear and non-linear regularizers require a value to control the intensity of the penalization. For our non-linear regularizer, this value is a parameter $\gamma\,\in\:]0, 1] \subset \mathbb{R}$ multiplied by the loss function's result.

\subsection{DL-SCA modular network loss function}
Now that we know the two losses required by our architecture as well as the hyperparameter to control their intensity, we have the expression (\ref{eq:SCA-CateAE_loss}) that defines the loss function for a DL-SCA modular network architecture.

\begin{equation}\label{eq:SCA-CateAE_loss}
\mathcal{L}_{\text{DL-SCA modular network}}=\gamma \cdot \mathcal{L}_{CE} + \omega \cdot \mathcal{L}_{\textrm{MSE}}
\end{equation}

Notice that there is an $\omega$ parameter for $\mathcal{L}_{\textrm{MSE}}$ that works exactly as $\gamma$. We fix $\omega=1$, because our goal is to control the regularization and not the reconstruction.

\subsection{Training strategy for a DL-SCA modular network}

Recently, authors from~\cite{paguada2021being} published an early stopping framework to monitor the state of a deep learning model during its training preventing it from getting overfit/underfit. Overfitting/underfitting is a phenomenon that might happen during training; it represents the state when a deep learning network cannot generalize beyond its training set.

The framework computes the guessing entropy at the end of each epoch basing the stopping criterion on the whole guessing entropy vector, considering when the guessing entropy converges, and how many traces keep the guessing entropy in the state of convergence, proving to overcome existing frameworks (more details can be found in the original paper~\cite{paguada2021being}). We use this early stopping to elaborate a training strategy for our DL-SCA modular network.

\paragraph{Training strategy} We know that an early stopping framework stops the training of a deep learning model when it meets conditions established using a metric, \textit{e.g.}, the accuracy of the model. Typically, these frameworks evaluate the entire model. In contrast, we need the framework to consider just the encoder and the classifier as they are the parts used in the SCA evaluation. The framework from~\cite{paguada2021being} is a ``typical'' framework, so it monitors the whole deep learning model.

We modified the suggested framework to receive a \textit{truncated model} which comprises just the modules of interest (encoder and classifier). To apply this modification is effortless when using the weight sharing technique~\cite{zhang2018learning}. In this technique two or more neural networks share the references to some specific layers, all of those networks can update the weights of those layers; however, in our particular case the original networks updates the weights, while the truncate model monitors the state (of those weights) of the encoder and classifier.

\begin{figure}[!ht]
\centering
\includegraphics[width=2.9in]{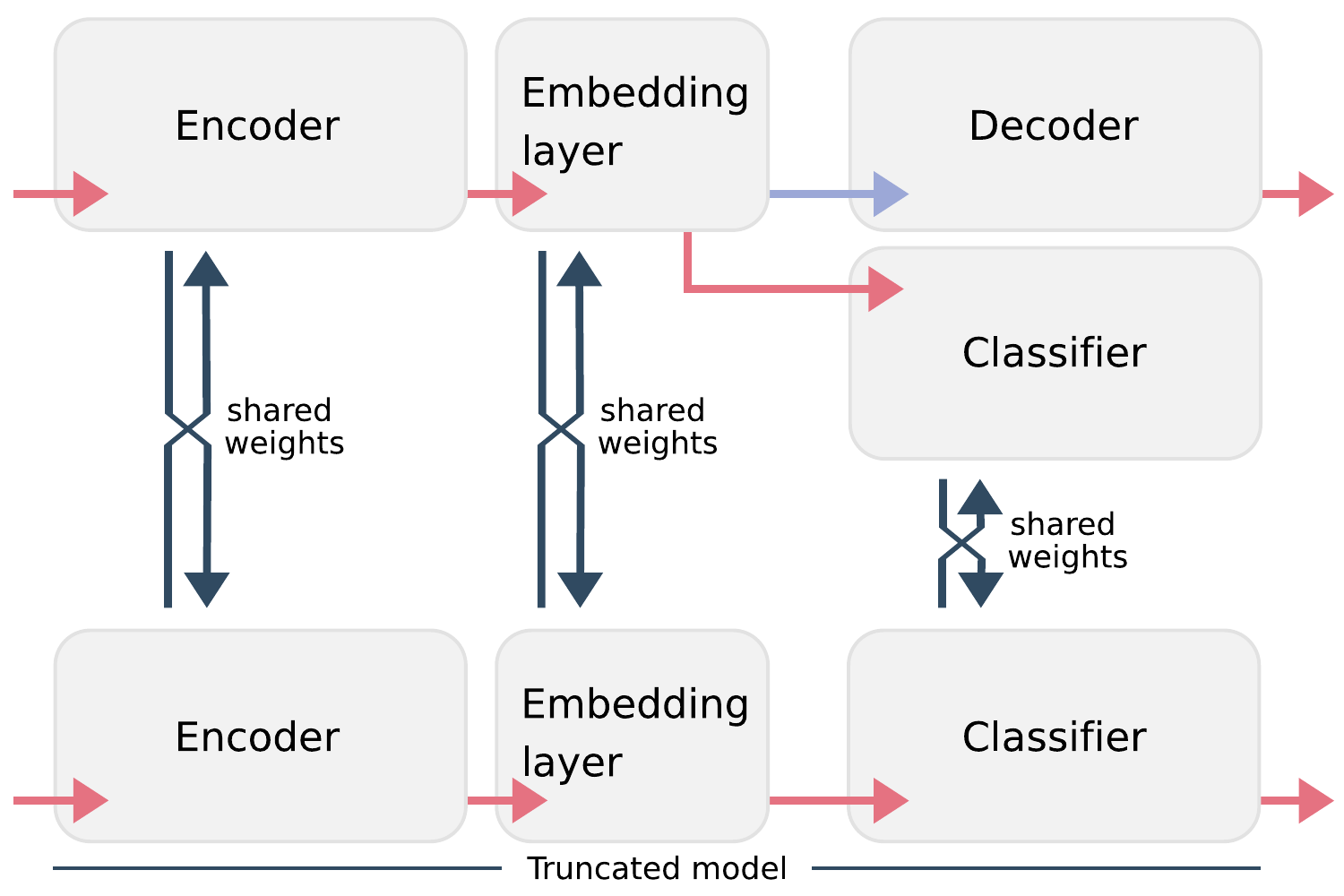}
\caption{The truncated model shares weights with the DL-SCA modular network; while the latter is training, the former updates his weights. The early stopping framework uses the truncated model to compute the guessing entropy at the end of each epoch, and it stops the training when it meets the conditions.}
\label{fig:shared_weights_truncate_model}
\end{figure}

Precisely, we set a truncated model (see Fig.~\ref{fig:shared_weights_truncate_model}) to reference those modules and to be only evaluated (not trained) by the early stopping framework. Then, it stops training when the encoder generates features that makes the classifier to achieve an expected performance, in this case an expected guessing entropy convergence. The modification works since the framework uses the truncated model as the predictor, and its output serves as the input to compute the guessing entropy. 

In the first experimental results section of this paper, we show the \textit{training strategy outcome} using surface plots. Notice that we did not stop the network training, so the surface plots correspond to the entire training process, our goal is to show that an SCA-DL modular network does not rely on an early stopping framework.

\section{Experimental results of training modules}
\label{sec:training_modules_exp_results}
In this section, we discuss the results of using our proposed approach over ASCAD datasets --- $\text{ASCAD}^f\:all\:desync$ and $\text{ASCAD}^r\:all\:desync$. 

We organize the experimental results as two different use cases where a DL-SCA modular network analyzes; (i) ASCAD fixed key dataset and (ii) ASCAD random key dataset. We accomplish two goals with these uses cases; (i) to show the feasibility when an SCA evaluation uses our architecture to attack an specific dataset, and (ii) to create a scenario where we demonstrate the feasibility of sharing modules. The strategy is applicable to real evaluations; the derived modular network evaluates a first dataset; consequently, a second modular network could evaluate another dataset borrowing a module from a previous modular network.

In our particular case, our experiments use two datasets that share the same source of data; precisely, both datasets were composed with leakage traces from the same microcontroller (Atmega8515 8-bit). We aim for performing experiments when the source of data is uncommon between both datasets as future works.

Notice, we used the same model for all levels of desynchronization, meaning that additional effort in finding neural network architectures for specific noisy scenarios is not required.

\subsection{$\text{ASCAD}^f$ \textit{all desync} use case}

The TABLE~\ref{tab:ascad_f_architecure} summarizes the hyperparameters of the modular network architecture to evaluate $\text{ASCAD}^f\:all\:desync$.

\subsubsection{Network's architecture} We set the architecture by following the discussion in Sect.~\ref{sec:autoencoder}; the first convolutional block uses dilated convolutions to avoid any useless features that might reduce the model's performance. We dilate the convolutions at the first convolutional block because it is where we deal with the original version of the trace. Further, we add convolutional blocks to the encoder following the rules applied for VGG~\cite{simonyan2014very} base deep learning architectures\footnote{VGG base architectures increase their number of kernels in convolutional layers by the power of 2}.

\begin{table}[!ht]
\renewcommand{\arraystretch}{1.0}
\label{tab:ascad_f_architecure}
\centering
\begin{tabular}{l r}
\hline
\multicolumn{2}{c}{\textbf{ENCODER}}\\
\hline
\textbf{Layer type} & \textbf{Details}\\
\rowcolor{Gray}
1. Conv & \# kernels: 32, kernel size: 64, SeLU\\
\rowcolor{Gray}
& dr: 3, kernel init: \textit{He\_Uniform}\\
\rowcolor{Gray}
2. Batch Norm & \\
\rowcolor{Gray}
3. Pooling & Average, \# kernels: 2, stride: 2\\
4. Conv & \# kernels: 64, kernel size: 25, SeLU\\
& kernel init: \textit{He\_Uniform}\\
5. Batch Norm & \\
6. Pooling & Average, \# kernels: 25, stride: 25\\
\rowcolor{Gray}
7. Conv & \# kernels: 128, kernel size: 3, SeLU\\
\rowcolor{Gray}
 & kernel init: \textit{He\_Uniform}\\
\rowcolor{Gray}
8. Batch Norm & \\
\rowcolor{Gray}
9. Pooling & Average, \# kernels: 5, stride: 5\\
10. Flatten & \\
11. Dense & 300 Units (Latent space)\\
\hline
\multicolumn{2}{c}{\textbf{DECODER}} \\
\hline
\textbf{Layer type} & \textbf{Details}\\
\rowcolor{Gray}
1. ConvTranspose & \# kernels: 128, kernel size: 3, SeLU\\
\rowcolor{Gray}
& stride: 7, kernel init: \textit{He\_Uniform}\\
\rowcolor{Gray}
2. Batch Norm &\\
3. ConvTranspose & \# kernels: 64, kernel size: 25, SeLU\\
& stride: 25, kernel init: \textit{He\_Uniform}\\
4. Batch Norm &\\
\rowcolor{Gray}
5. ConvTranspose & \# kernels: 32, kernel size: 64, SeLU\\
\rowcolor{Gray}
& stride: 2, kernel init: \textit{He\_Uniform}\\
\rowcolor{Gray}
6. Batch Norm &\\
7. ConvTranspose & \# kernels: 1, kernel size: 1, Sigmoid\\
& stride: 1, kernel init: \textit{He\_Uniform}\\
\hline
\multicolumn{2}{c}{\textbf{CLASSIFIER}}\\
\hline
\textbf{Layer type} & \textbf{Details}\\
\rowcolor{Gray}
1. Conv & \# kernels: 4, kernel size: 1, SeLU\\
\rowcolor{Gray}
& dr: 3, kernel init: \textit{He\_Uniform}\\
\rowcolor{Gray}
2. Batch Norm & \\
\rowcolor{Gray}
3. Pooling & Average, \# kernels: 2, stride: 2\\
\rowcolor{Gray}
4. Flatten &\\
5. Dense & 10 (units), SeLU\\
6. Dense & 10 (units), SeLU\\
7. Dense & 10 (units), SeLU\\
8. Dense & 256 (units), Softmax\\
\hline
\end{tabular}
\caption{DL-SCA modular network architecture to use in experiments with $\text{ASCAD}^f$ all desynchronizations levels}
\end{table}

The decoder mirrors the encoder, as our downsampler uses symmetric autoencoders. For the decoder to up-sample, namely to reconstruct the actual length of the trace, it uses transpose convolutions. As known, matrix multiplication is not commutative, and we cannot achieve the same output in respective convolutional blocks. Consequently, we have to tune the hyperparameters in the decoder's convolution layers. For instance, let us take the third encoder's convolutional block that uses a stride value of $5$, its corresponding decoder's transpose convolutional block is the first one but it uses stride value of $7$. By doing this, we fix the output of the decoder to meet the original trace dimension.

\begin{figure}[!ht]
\centering
\includegraphics[width=2.5in]{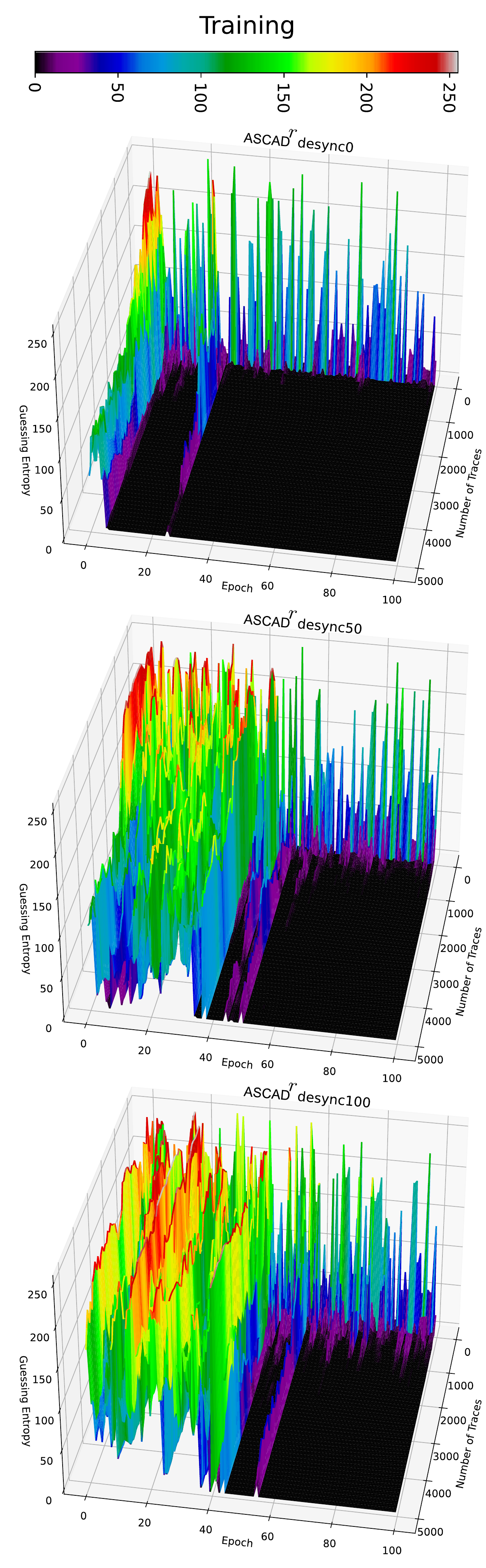}
\caption{The training process of the modular network for $\text{ASCAD}^f$ datasets. The surface represents the values of the guessing entropy during a chosen number of epochs. Stopping condition success GE=0.}
\label{fig:ge_ascad_f_epochs}
\end{figure}

\subsubsection{Latent space hyperparameters}
With regard to latent space units and $\gamma$ value. We perform a grid search for the best number of units in the latent space, using the values 100, 200, 300, 400, and 560. Further, we know that the parameter $\gamma$ relates strictly with the number of latent units; consequently, to find the value of $\gamma$ we create combinations using the latent space values and values of $\gamma$ as $\{1e^{-3}, 1e^{-6}, 1e^{-9}\}$. It turned out that the best combinations was 300, and $1e^{-3}$ for latent space units and the $\gamma$ parameter, respectively.

Regarding the classifier module, bear in mind that we are only interested in its classification performance and not too much in its ability to filter out unnecessary features of the leakage traces, so we use a shallow architecture since it will deal with already filtered features.

\subsubsection{Training strategy and results} 
\begin{figure}[!ht]
\centering
\includegraphics[width=2.5in]{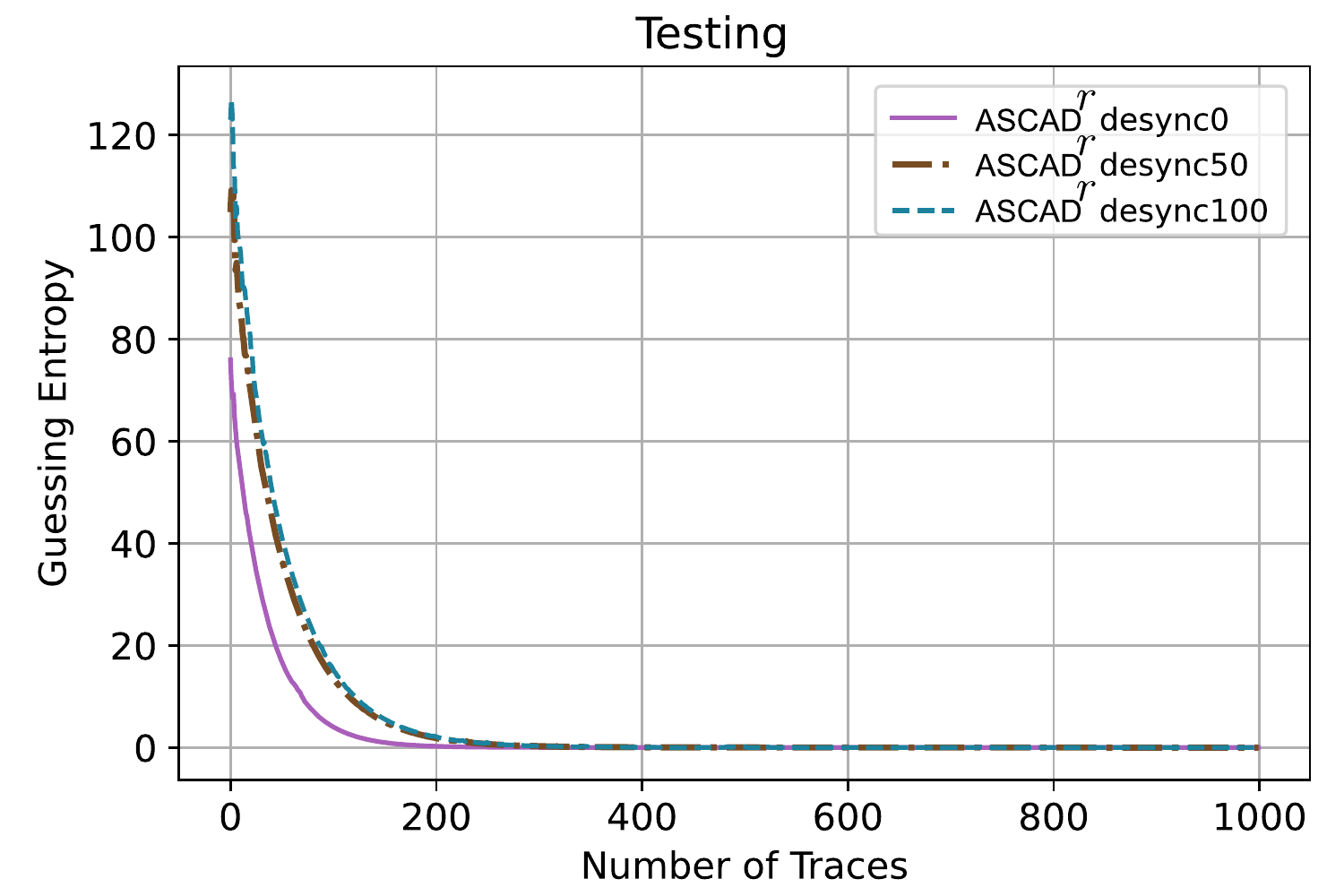}
\caption{Guessing entropy over $\text{ASCAD}^f$ all levels of desynchronization}
\label{fig:ge_ascad_f_dataset}
\end{figure}

As we said, to train a modular network, we use the early stopping framework from~\cite{paguada2021being}. To show that our suggested architecture does not rely on the framework, we did not stop the training after the mentioned framework finds the best learning state. Further, we will use this outcome in the next section to discuss the result of the reusing modules experiment. Fig.~\ref{fig:ge_ascad_f_epochs} depicts the training process when our modular network evaluates $\text{ASCAD}^f$ datasets. As we expected, the training outcome differs according to the level of desynchronization; regardless, our modular network achieved a zero convergent guessing entropy for all desynchronization levels. A view of the attack performance is depicted in Fig.~\ref{fig:ge_ascad_f_dataset}.

\subsection{$\text{ASCAD}^r$ \textit{all desync} use case}

\begin{figure}[!ht]
\centering
\includegraphics[width=2.5in]{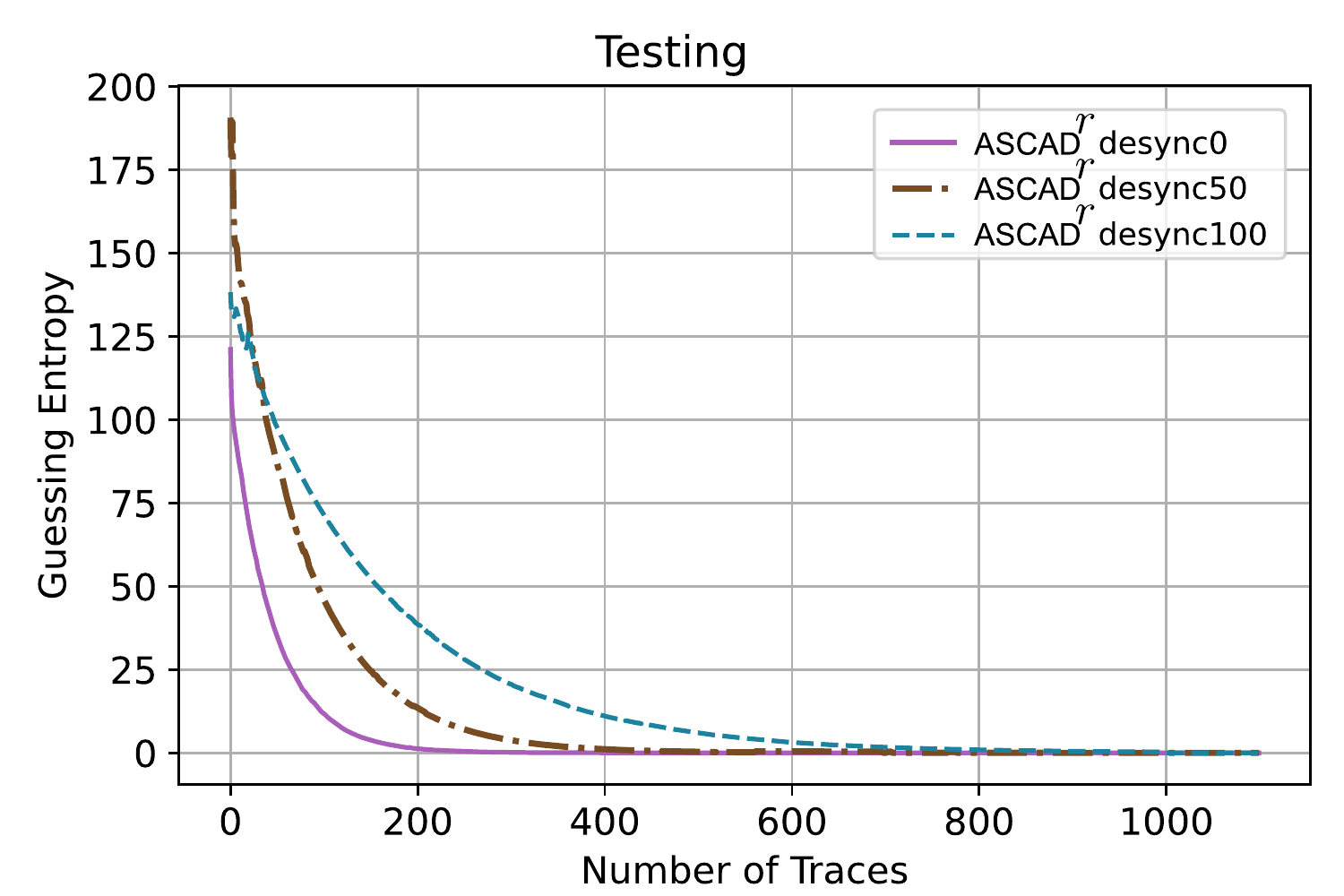}
\caption{Guessing entropy over $\text{ASCAD}^r$ all levels of desynchronization}
\label{fig:ge_ascad_r_dataset}
\end{figure}

\begin{figure}[!ht]
\centering
\includegraphics[width=2.5in]{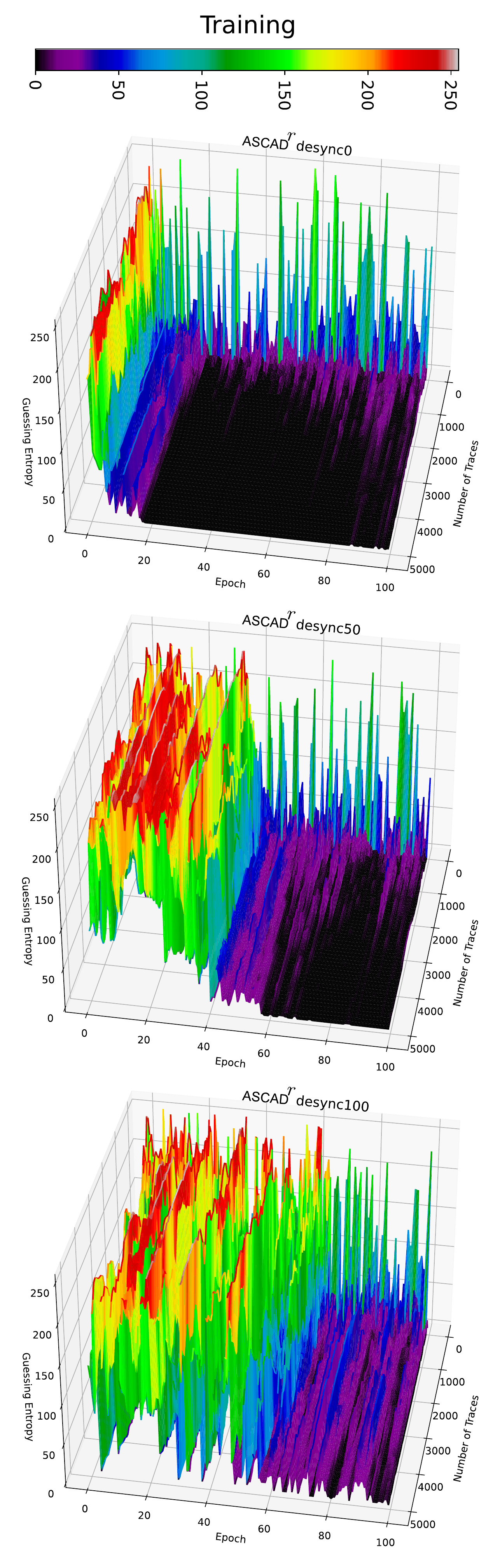}
\caption{The training process of the modular network for $\text{ASCAD}^r$ datasets. The surface represents the values of the guessing entropy during a chosen number of epochs.}
\label{fig:ge_ascad_r_epochs}
\end{figure}

\subsubsection{Network's architecture} Regarding this dataset, our strategy was to keep the same modular network as the previous use case to reuse as much as possible an already worked model and see how it performs. After experimenting, we noticed that the downsampler module required an additional convolutional block ---identical to the third convolutional block of the decoder--- without pooling layer. Consequently, the decoder should also have the corresponding transpose convolutional block.

\subsubsection{Latent space hyperparameters, training strategy, and results}
We keep the same classifier as in the previous use case because we have the same number of latent units. In our particular case, to keep the same latent units is convenient because we aim for exchanging a trained classifier in the following experiments to evaluate the modular re-usability. Fig.~\ref{fig:ge_ascad_r_epochs} depicts the training process of guessing entropy by epochs for $\text{ASCAD}^r$ dataset. In this case, we observe that the performance of our modular network slightly decreases, which is expected since the dataset has a higher level of noise than the previous. Even though, we achieve good guessing entropy convergence as depicts Fig.~\ref{fig:ge_ascad_r_dataset}. Finally, we compare our experimental results with previously reported results over the same datasets. TABLE~\ref{tab:ascad_comparison_values} gathers this information.

\begin{table}[!ht]
\renewcommand{\arraystretch}{1.3}
\newcolumntype{a}{>{\columncolor{Gray}}c}
\newcolumntype{n}{>{\columncolor{Gray}}l}
\centering
\begin{tabular}{ncccaaa}
\hline
& \multicolumn{3}{c}{\bfseries {$\text{ASCAD}^f$}} & \multicolumn{3}{a}{\bfseries $\text{ASCAD}^r$}\\
& \multicolumn{3}{c}{desync level} & \multicolumn{3}{a}{desync level}\\
\multirow{-3}{*}{\textbf{Reference}}& 0 & 50 & 100 & 0 & 50 & 100 \\
\hline
From~\cite{paguada2021towardAE} & 791 & - & - & - & - & 1500 \\
From~\cite{paguada2020ForgottenHy} & 178 & - & - & 550 & - & 3\,000 \\
From~\cite{Zaid2019} & 191 & 244 & 270 & - & - & - \\
\textbf{This work} & 148 & 231 & 236 & 210 & 407 & 1092 \\
\hline
\end{tabular}
\caption{Best guessing entropy convergence comparison}
\label{tab:ascad_comparison_values}
\end{table}

\section{Module re-usability experimental results}
\label{sec:module_re-usability_exp_results}
This section presents the results of module re-usability. We show that another non-trained DL-SCA modular network can reuse the modules of a DL-SCA modular network. We use the DL-SCA-based module networks trained in the previous section to prove it.

\subsection{Analyzing transferability}
We aim to show how ``transferable'' is the knowledge of a classifier module. We have six modular network ---meaning six classifiers--- trained with three different datasets ---three on $\text{ASCAD}^f$ and three on $\text{ASCAD}^r$. Further, due to the number of latent units (300) we used, all classifiers are interchangeable without performing additional downsampling operations to fix their inputs. For our experiments, we took the classifier from the DL-SCA modular network of $\text{ASCAD}^f\:desync50$ to share with all the downsampler from $\text{ASCAD}^r$. We considered it sufficient for proving our claim about ``module re-usability''. We chose $\text{ASCAD}^f \longmapsto \text{ASCAD}^r$ direction because it represents the complex direction ---from fixed key to random key.

\begin{table}[!ht]
\renewcommand{\arraystretch}{1.2}
\newcolumntype{a}{>{\columncolor{Gray}}m{2.5cm}}
\newcolumntype{b}{>{\columncolor{Gray}}m{3cm}}
\centering
\begin{tabular}{bla}
\hline
& \textbf{Original}  & \textbf{Gradient} \\
\multirow{-2}{*}{\textbf{Shared classifier}} & \textbf{downsampler \& classifier}& \textbf{operation}\\
\hline
& \multirow{2}{*}{$\text{ASCAD}^r\:desync0$} & heatmap\\
& & gradient vis\\
& \multirow{2}{*}{$\text{ASCAD}^r\:desync50$} & heatmap\\
& & gradient vis\\
& \multirow{2}{*}{$\text{ASCAD}^r\:desync100$} & heatmap\\
\multirow{-6}{*}{$\text{ASCAD}^f\:desync50$}& & gradient vis\\
\hline
\end{tabular}
\caption{Summary of the similarity analysis between $\text{ASCAD}^f\:desync50$ classifier and $\text{ASCAD}^r\:all\:desync$ classifiers}
\label{tab:heatmaps_gradvis_combinations}
\end{table}

We inspect the transferability of the $\text{ASCAD}^f\:desync50$ classifier by conducting a similarity analysis using gradient activation operations. In particular, we use heatmaps and gradient visualization to compare how the neurons' of the classifier are activated by the data outputted from the downsamplers.

We perform this analysis by locking specific layers of the classifier to identify how transferable those layers are. Precisely, we choose convolutional block (Conv) layers and fully connected block (FC) layers and lock them by turns to evaluate them separately. A heatmap allows us to inspect the convolutional layers of the classifier, while gradient visualization helps us analyze how both Conv and FC perform with the different datasets.

\begin{figure}[!ht]
\centering
\includegraphics[width=2.5in]{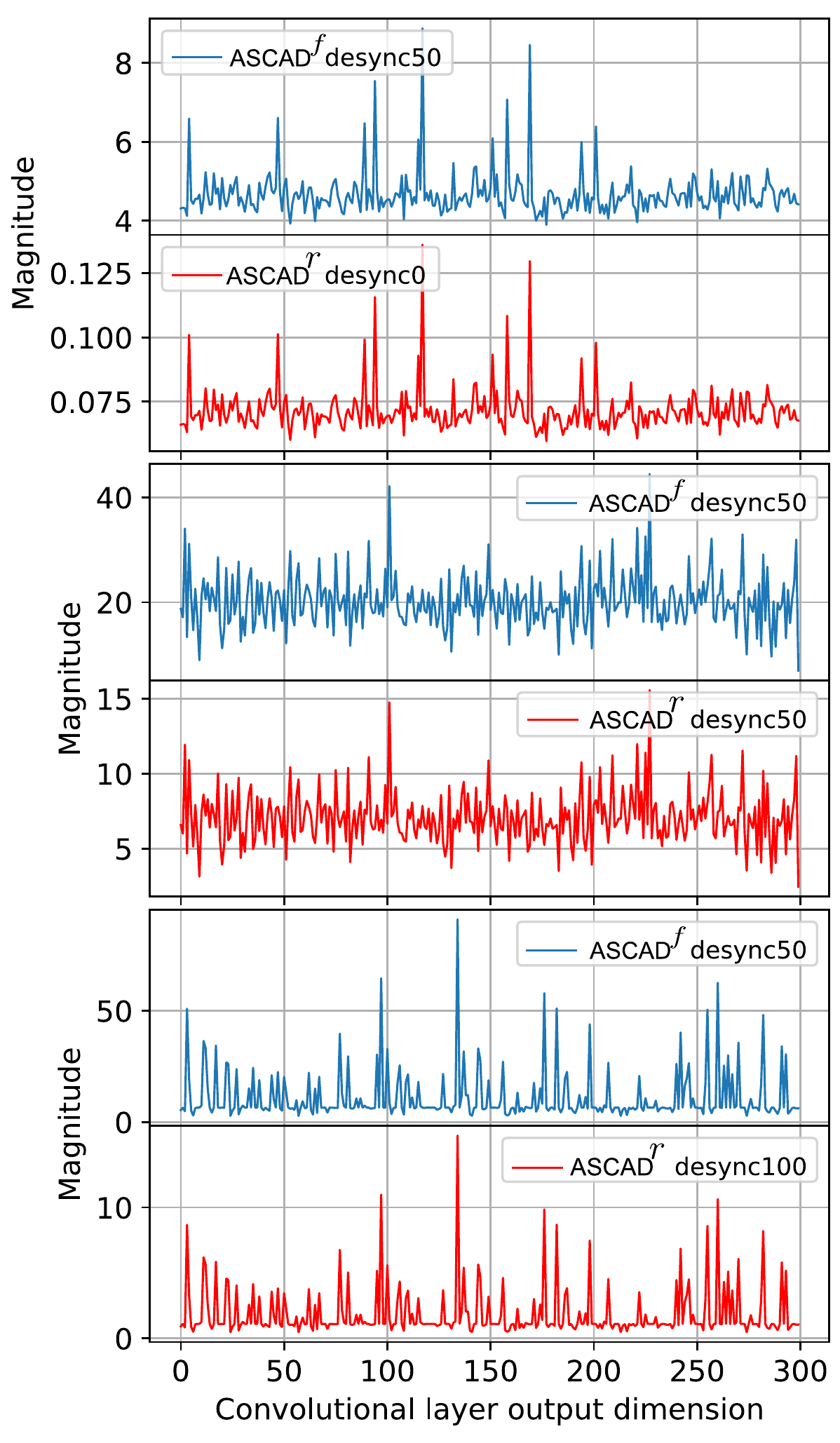}
\caption{Comparison between heatmaps of the $\text{ASCAD}^f\:desync50$ classifier and classifiers from all the $\text{ASCAD}^r$ datasets. Notice how $\text{ASCAD}^f\:desync50$ heatmap resembles all other heatmaps. It indicates that $\text{ASCAD}^f\:desync50$ classifier's convolutional layer fires its neurons according the data received.}
\label{fig:hetmap_classy50}
\end{figure}

TABLE~\ref{tab:heatmaps_gradvis_combinations} summarizes the similarity analysis we are going to perform using the classifier $\text{ASCAD}^f\:desync50$, the $\text{ASCAD}^r$ datasets, and the gradient activation operations. Fig.~\ref{fig:hetmap_classy50} depicts the first convolutional layer heatmaps from $\text{ASCAD}^f\:desync50$ classifier and $\text{ASCAD}^r\:all\:desync$ classifiers ($desync0$, $desync50$, and $desync100$).

For these particular experiments, all $\text{ASCAD}^r$ classifiers share similarities with the $\text{ASCAD}^f\:desync50$ classifier in how their convolutional layer neurons' get stimulated. According to our assumptions, it indicates that the weights of those layers might be transferable. This claim is experimentally demonstrated later in the final experiments.

Although the magnitude of the $\text{ASCAD}^f\:desync50$ classifier's heatmap is higher than any other heatmap from $\text{ASCAD}^r$ classifiers, it does not represent a drawback to the transferability. We could have gotten the same magnitudes if we had normalized the weights applying constraints in the architecture, though a similarity analysis does not need to do this.

\begin{figure}[!ht]
\centering
\includegraphics[width=2.5in]{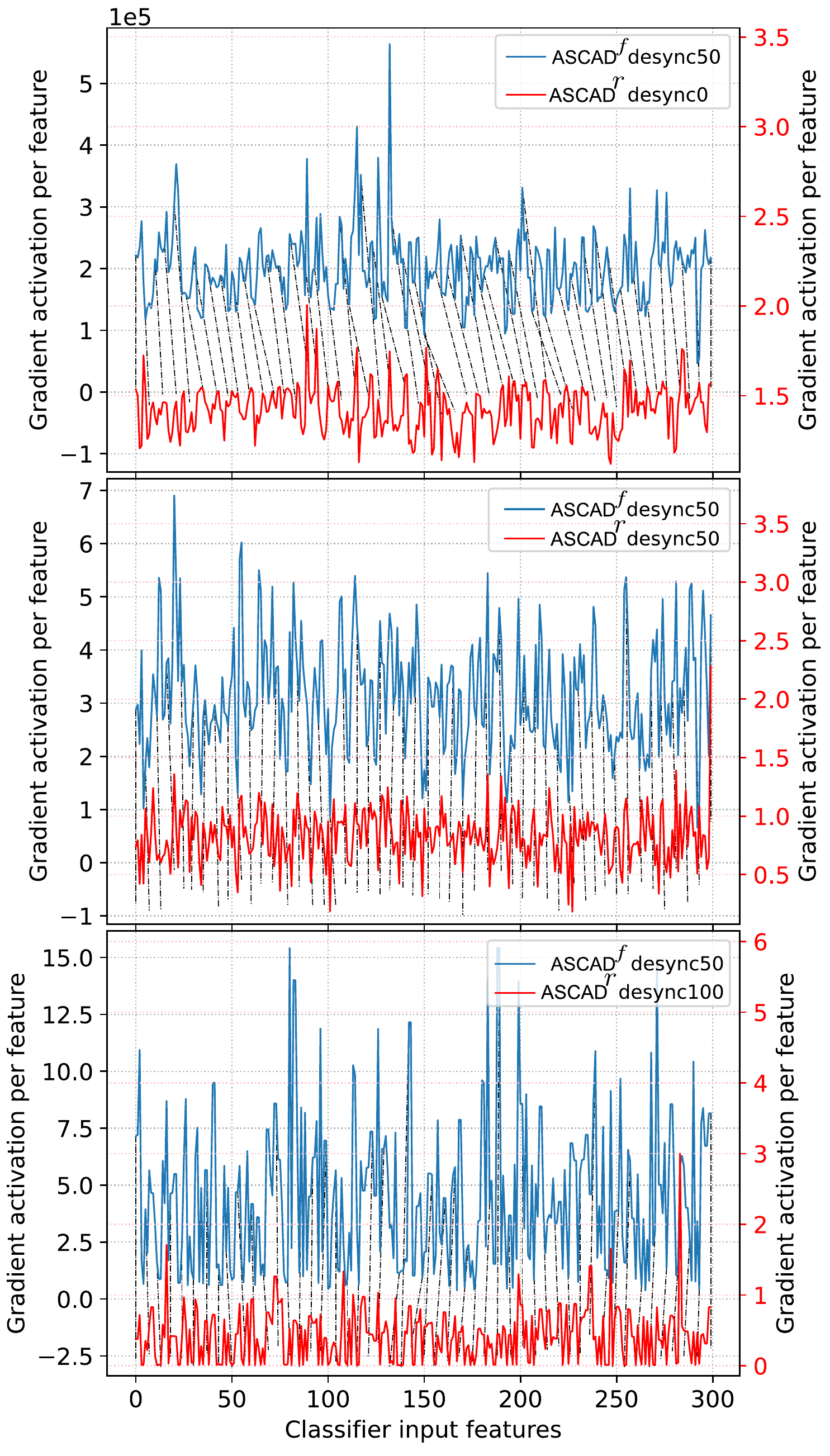}
\caption{Comparison between gradient activation per sample of the $\text{ASCAD}^f\:desync50$ classifier and classifiers from all the $\text{ASCAD}^r\:all\:desync$.}
\label{fig:gradvis_classy50}
\end{figure}

We use gradient visualization to inspect the classifiers' fully connected block (FC). The output of that operation indicates which input \textit{features} are the most \textit{meaningful} for the classification. The gradient visualization uses the loss function of a trained classifier to conduct backpropagation, collecting the information about those neurons that emphasize the performance. Further, when it reaches the input layer, it points out which features are connected to those neurons, indicating the meaningful features~\cite{masure2019gradient,shrikumar2016not,ancona2017towards}. Fig.~\ref{fig:gradvis_classy50} depicts the result of gradient visualization operation.

Notice that gradient visualization shows less intuition than heatmap. As a workaround, we apply a Dynamic Time Warping (DTW)~\cite{muller2007dynamic} to visualize the similarities between gradient visualization signals. 

According to this experiment, two phenomena happen; (i) the meaningful features are displaced according to each classifier, or/and (ii) the meaningful features are less intense in magnitude. These phenomena could represent an issue. For instance, let us take the $\text{ASCAD}^r\:desync0$ classifier, notice the displacement because the $\text{ASCAD}^f\:desync50$ interprets that the meaningful features localize differently. Further, those features have an even lower magnitude in contrast to those supposedly being the lowest(see points from 0 to 30 in Fig.\ref{fig:gradvis_classy50} top plot).

This analysis gives us the intuition that we will need to retrain the classifier; nevertheless, the reader might remember that the classifier is just a part of a bigger model. The downsampler will leverage its learning according to the limitation imposed by the classifier.

\subsection{Playing with blocks}
Let us suppose we have trained a DL-SCA modular network using a former dataset; then, we have the opportunity to evaluate another dataset. We could use the classifier module of the first network to evaluate it. In this hypothetical scenario, the first dataset is played by $\text{ASCAD}^f$ and the second one by the $\text{ASCAD}^r$ dataset. 

To experimentally evaluate if we need to re-train some or all parts of the classifier, we perform experiments locking the blocks of the classifier to restrict them from getting trained. In the previous sub-section, we inspected the blocks of the classifier (Conv and FC), and we observed some similarities in its neurons' weights. Now, we are going to evaluate the performance of the whole modular network when its classifier module has the following locks:

\begin{itemize}
\item Convolutional block
\item Fully-connected block
\item Both blocks
\end{itemize}

We will refer to these as ``sharing protocols''. We find out which could be the best sharing protocol for these particular modular networks by locking the blocks. TABLE~\ref{tab:re-sharing_modules} summarizes the combination of locks and dataset where the shared classifier will be used. 

\begin{table}[!ht]
\renewcommand{\arraystretch}{1.2}
\newcolumntype{a}{>{\columncolor{Gray}}m{3.5cm}}
\newcolumntype{n}{>{\columncolor{Gray}}m{3cm}}
\centering
\begin{tabular}{anl}
\hline
\multicolumn{2}{c}{\cellcolor{Gray}\textbf{Classifier}} & \textbf{Re-used in} \\
\textbf{Shared classifier} & \textbf{Lock use case} & \textbf{Dataset} \& \textbf{filter}\\
\hline
& Convlock & \multirow{3}{*}{$\text{ASCAD}^r desync0$}\\
& Bothlock &\\
& FClock &\\

& Convlock & \multirow{3}{*}{$\text{ASCAD}^r desync50$}\\
& Bothlock &\\
& FClock &\\

& Convlock & \multirow{3}{*}{$\text{ASCAD}^r desync100$}\\
& Bothlock & \\
\multirow{-9}{*}{$\text{ASCAD}^f\:desync50$}& FClock &\\
\hline
\end{tabular}
\caption{Combination of sharing protocols used for the $\text{ASCAD}^f\:desync50$ classifier}
\label{tab:re-sharing_modules}
\end{table}

We previously said that the chosen classifier $\text{ASCAD}^f\:desync50$ will tackle a more complex dataset ---the $\text{ASCAD}^r\:desync100$. Now, by evaluating the $\text{ASCAD}^r\:desync0$ dataset; then, we will cover the scenario where the shared classifier comes from a more complex dataset. Still, bear in mind that it is in terms of desynchronization because it does not come from a complex dataset in terms of its secret key's nature ---from random to fixed key, for example. So, we rate the ``experience'' of the classifier as \textit{medium level of experience}.

\begin{figure*}[!ht]
\centering
\includegraphics[width=0.98\textwidth]{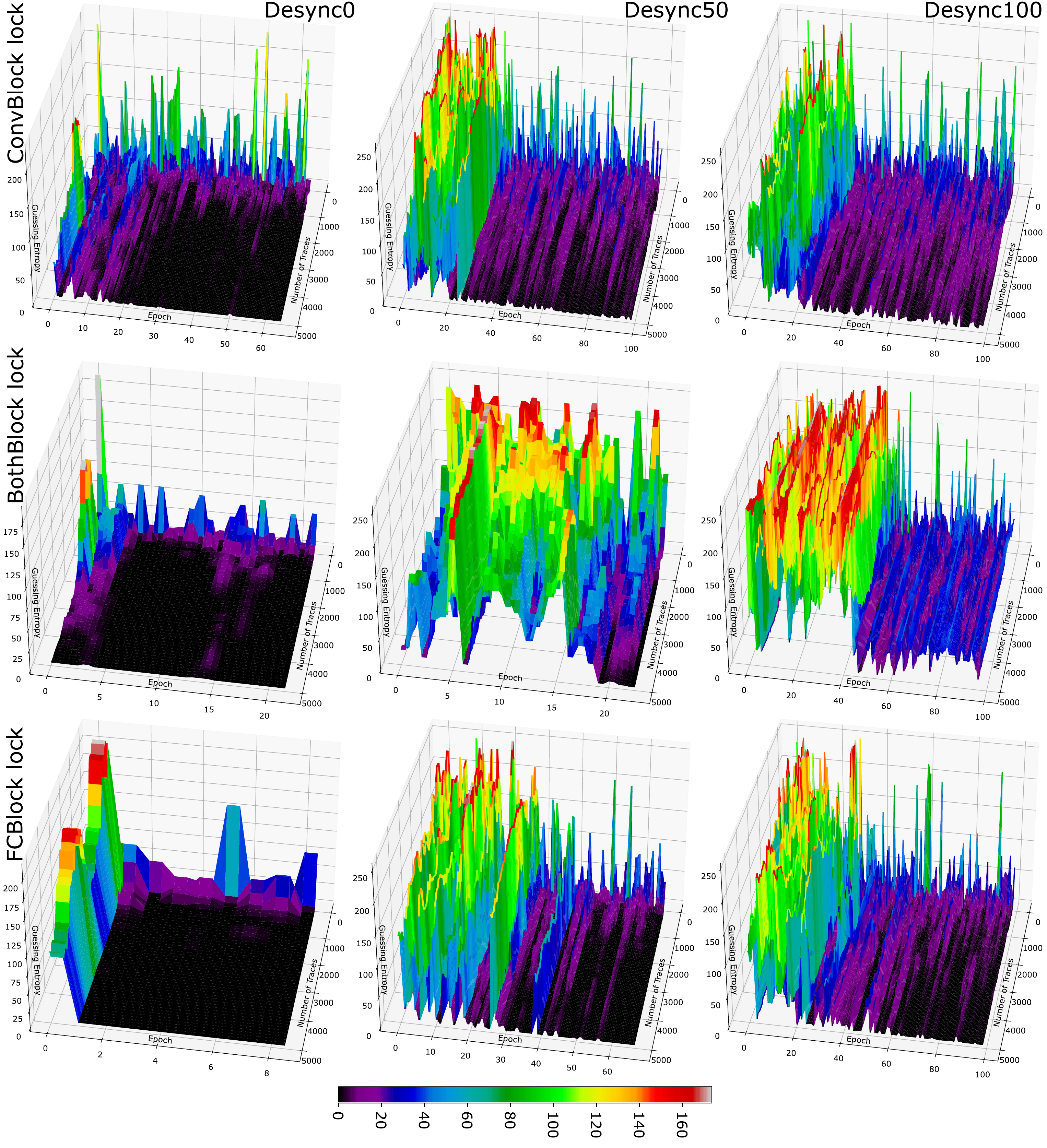}
\caption{The training results of the knowledge transferability experiments. Through the columns lies the levels of desynctronization [0, 50, 100]; while through the rows, lies the different block lock cases ---ConvLock, BothLock, and FCLock.}
\label{fig:all_enhancing}
\end{figure*}

Due to space constraints, we did not perform an inter-classifier sharing and a no-block lock sharing protocol; furthermore, we claim that the sharings addressed in our experiments represent the difficult one, being enough to prove our contribution. However, we let those experiments and further combinations of sharing protocol for future works. Fig.~\ref{fig:all_enhancing} depicts the training process of all chosen sharing protocols. It is worthy of mentioning that we did not change the loss intensity parameter ($\gamma$), reducing the effort in tuning the modular network. 

For this experiments, we trained the modular networks using the early stopping framework from~\cite{paguada2021being}. Contrary we did in the previous section, we do stop the training when the policy finds out the best learning state. We can now know the number of epochs required to achieve good performance.

\begin{figure}[!ht]
\centering
\includegraphics[width=2.5in]{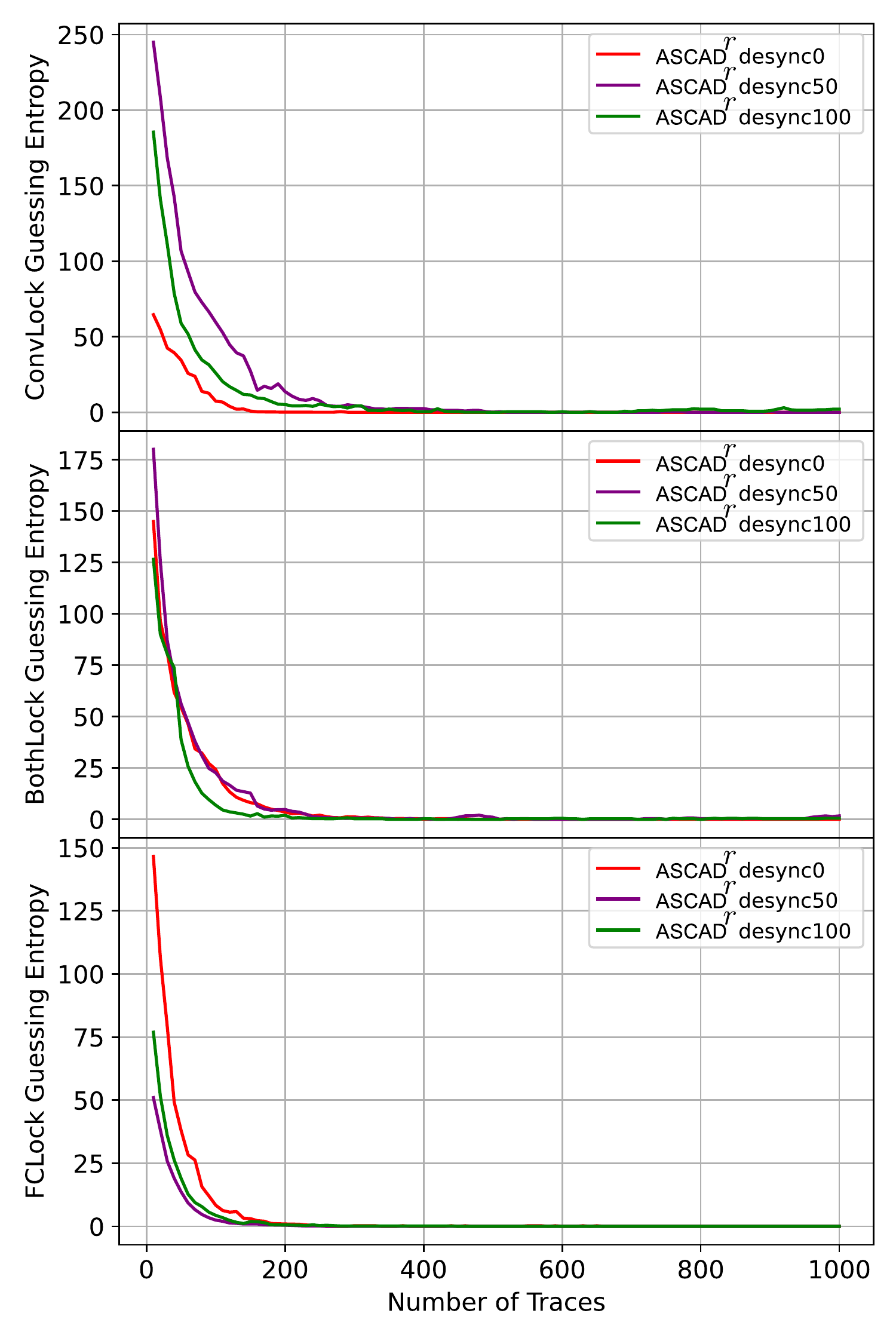}
\caption{Best guessing entropy results from all sharing protocols.}
\label{fig:all_ge_transfer}
\end{figure}

Generally, all sharing protocols perform well if we contrast the training process of Fig.~\ref{fig:all_enhancing} and Fig.~\ref{fig:ge_ascad_f_dataset}. Nevertheless, the sharing protocols that worked best are the \textit{fully-connected block}, \textit{both blocks}, and the \textit{convolutional block lock}. 

Observe that for the fully-connected block lock, $\text{ASCAD}^r\:desync0$ has a convergent guessing entropy after 9 epochs, $\text{ASCAD}^r\:desync50$ at 65 epochs, and $\text{ASCAD}^r\:desync100$ took the whole training process (100 epochs); even thought, it achieves good performance. Both blocks lock cases seem to require more epoch or the convergence is roughly achieved, $\text{ASCAD}^r\:desync0$ and $\text{ASCAD}^r\:desync50$, for instance. Finally, we notice that the convolutional block lock converges after several more epochs than the previous locks. In this case, $\text{ASCAD}^r\:desync100$ did not converge within $1\,000$ leakage traces. We summarize in Fig.~\ref{fig:all_ge_transfer} the best guessing entropy from all combination of locks.

\subsection{Discussion}
Using a shared classifier instead of a non-trained modular network, we have reduced the training time and the effort in tuning hyperparameters while evaluating the leakage of a dataset with good results. Since we locked some blocks and the whole classifier, we reduced the number of neurons to train; consequently, the training time is reduced since the number of operations per neuron is less than a non-trained modular network. As we do not have to tune the hyperparameter of a classifier, then we do not spend time in it. Further, we are confident that the classifier has a high probability of working since it already has previous ``experience''. We demonstrated the latter by actually achieving good results.

Clearly, some initial effort has to be made. For instance, we were tuning the latent space and losses intensity hyperparameters. Coming up with an initial deep learning modular network could be challenging, but it is an equivalent effort in finding several small deep learning models for different datasets. Finally, bear in mind that by saying that a classifier has previous experience, we do not claim that it will work flawlessly. As we said, the experience of a shared classifier represents a neurons' weights initializer. So instead of randomly initializing the weights using any well-known function ---he uniform, for instance---; we start from a state leveraged by a previous worked learning. We have proved, experimentally, that it has good results.

\section{Conclusion}
\label{sec:conclutions}
We introduced the DL-SCA modular network approach to conducting SCA evaluation reusing modules from previously trained modular networks. A DL-SCA modular network consists of two main modules; a downsampler and a classifier. We demonstrate that modules from a modular network can be detached and attached to other modular networks and conduct an efficient SCA evaluation. The strategy is to use a classifier with good performance and reuse it to conduct another evaluation in a different dataset.

Our experiments demonstrate that it is not mandatory to re-train a classifier module to effectively evaluate the aimed dataset, regardless of whether the source classifier has been trained with a dataset with a lower noise level. We systematically lock the layers of the classifier to restrict them from getting trained, replicating different sharing protocols to evaluate the effectiveness of our approach.

As we said in the paper, we aim to work with more sharing protocols and improve the performance of our modular network in future works by using other types of deep learning architecture for the downsampler. Furthermore, we look for applying methodologies that might help tun the hyperparameters of a modular network.


%



\section*{Acknowledgment}
Acknowledge might change once this paper gets accepted.




\bibliographystyle{splncs04}
\bibliography{bib}
\end{document}